\UseRawInputEncoding
\documentclass{optica-article}
\journal{opticajournal} 
\usepackage{cancel}
\definecolor{Czer}{RGB}{200,46,46}
\usepackage{ulem,xpatch}

\articletype{Research Article}
\usepackage{lineno}
\xpatchcmd{\sout}
  {\bgroup}
  {\bgroup}
  {}{}

\newcommand\scalemath[2]{\scalebox{#1}{\mbox{\ensuremath{\displaystyle #2}}}}
\begin{document}

\title{Natural exceptional points in the excitation spectrum of a light-matter system}

\author{A.~Opala,\authormark{1,2,*}
M.~Furman,\authormark{1}
M.~Kr\'ol,\authormark{1}
{R.~Mirek},\authormark{1}
{K.~Tyszka},\authormark{1}
{B.~Seredy\'nski},\authormark{1}
{W.~Pacuski},\authormark{1}
{J.~Szczytko},\authormark{1}
{M.~Matuszewski},\authormark{2}
{B.~Pi\k{e}tka},\authormark{1}}

\address{\authormark{1}Institute of Experimental Physics, Faculty of Physics, \\University of Warsaw, ul. Pasteura 5, PL-02-093 Warsaw, Poland\\
\authormark{2}Institute of Physics, Polish Academy of Sciences,\\ Aleja Lotnik\'ow 32/46, PL-02-668 Warsaw, Poland}

\email{\authormark{*}aopala@fuw.edu.pl} 


\begin{abstract*} 
We observe natural exceptional points in the excitation spectrum of an exciton-polariton system by optically tuning the light-matter interactions. The observed exceptional points do not require any spatial or polarization degrees of freedom and result solely from the transition from  weak to strong light-matter coupling. We demonstrate that they do not coincide with the threshold for photon lasing, confirming previous theoretical predictions~\cite{Littlewood2019,Khurgin2020}. Using a technique where a strong coherent laser pump induces up-converted excitations, we encircle the exceptional point in the parameter space of coupling strength and  particle momentum. Our method of local optical control of light-matter coupling paves the way to investigation of fundamental phenomena including dissipative phase transitions and non-Hermitian topological states.

\end{abstract*}

\section{Introduction}
Understanding quantum systems coupled to the environment is one of the most important directions of research in quantum physics~\cite{breuer2016colloquium,rivas2012open}. In recent years, it became clear that open character of a quantum system may be viewed not only as a detrimental factor, but also as a resource that can be exploited for its control, and lead to observations of novel physical phenomena~\cite{Devoret_PhysRevA.88.023849,Verstraete2009}. Under certain assumptions, taking into account the effect of the environment leads to the description in terms of non-Hermitian Hamiltonians, which include dissipative terms~\cite{bender2007making}. 

One of the fundamental features of such systems is the possibility of the existence of exceptional points, which are points in  parameter space where eigenvalues and eigenvectors coalesce~\cite{Heiss2012}. These degeneracies possess a number of intriguing and non-intuitive properties. Adiabatic encircling of an exceptional point along a closed path in parameter space does not lead to the same eigenstate, but to exchange of eigenstates~\cite{Mohammad2019}. This fundamental property does not have a correspondence in Hermitian degeneracies such as Dirac points. Moreover, exceptional points may give rise to dissipative phase transitions~\cite{Weitz2021} and the appearance of genuinely non-Hermitian topological states~\cite{Szameit2015,Ueda_PhysRevX.8.031079}. On the practical side, it was suggested that the nonlinear dependence of eigenstate energies in the vicinity of an exceptional point may be used to design highly efficient sensors~\cite{Hodaei2017,Wiersig2020}. 

Thanks to the fundamental similarity of quantum mechanical and optical systems, exceptional points were realized experimentally in several systems with engineered optical gain and loss~\cite{Mohammad2019}. In particular, parity-time (PT) symmetric systems were designed for observations of the transition from a symmetric to a symmetry-broken state via crossing an exceptional point~\cite{Christodoulides_PhysRevLett.103.093902,Doiron}. Purposely designed systems with coupled optical modes, either confined in waveguides~\cite{Feng2017}, ring resonators~\cite{Chen2017,Yang2014} or traps~\cite{Gao2021} or spatially extended degenerate modes~\cite{Gao2015,Gao2018,Zhang2017} were used to realize non-Hermitian physics of fundamental two-mode coupling. 

Recently, it was suggested in several works that natural exceptional points may appear in systems where light and electron-hole excitations in semiconductors become strongly coupled, leading to the appearance of hybridized modes called exciton-polaritons~\cite{Littlewood2019,Khurgin2020}. We call these exceptional points ``natural'' since they do not require engineering of decay rates or polarization splitting, and appear naturally in any system at the transition from strong to weak coupling. In this case the light and matter modes play the role of coupled modes with mode-dependent gain and loss. In the regime of weak pumping, they interact strongly via vacuum Rabi coupling, which gives rise to mixed polariton modes. Upon decreasing the strength of coupling, the system goes through an exceptional point which marks the transition to weak coupling, where light and matter modes behave separately.  It was suggested that in exciton-polariton systems this dissipative transition may occur above the bosonic condensation threshold but below the photon lasing threshold, calling for the reinterpretation of experiments where the so-called ``second threshold'' was observed~\cite{Littlewood2019}. Experimentally, exceptional points were observed in exciton-polariton systems by manipulating the polarization degree of freedom~\cite{Kono_gao2018continuous,Krol_AnnihilationEP,Liao_EP,su2021direct}, in a chaotic non-Hermitian billard~\cite{Gao2015} and in a double well trap~\cite{Gao2021}. However, these experiments did not investigate natural exceptional points, since in all cases additional degrees of freedom, either spatial or corresponding to light polarization, were employed.

\begin{figure*}
\centering
  \includegraphics[width=\textwidth]{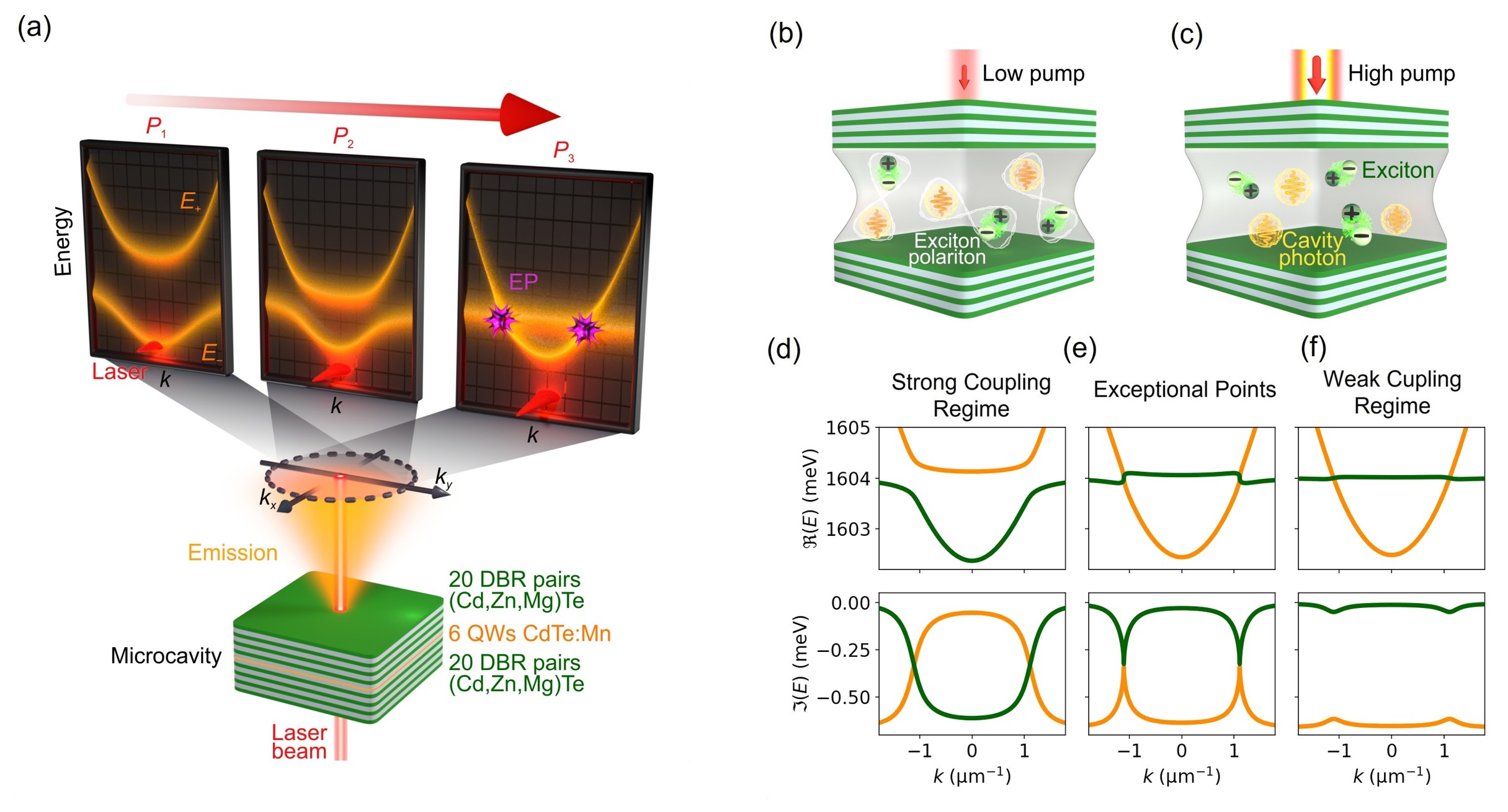}

  \caption{Scheme of the experiment. (a) The incident laser beam creates up-converted
 excitations with non-zero wave vector. With increasing pump power $P$, the coupling strength decreases, allowing observation of exceptional points (EP). Increase of resonant pumping intensity results in the transition from strong (b) to weak coupling (c) due to interaction with the light-induced incoherent particle reservoir. (d--f) Real (top) and imaginary (bottom) parts of eigenvalues of Hamiltonian~(\ref{eq:model}), showing the transition from spectra characteristic for strong coupling (polariton anticrossing) to weak coupling (simple crossing of exciton  and photon modes). Parameters are: $E_C^0=1602.5$ meV, $E_X^0=1604.0$ meV, $\gamma_C=1$ ps$^{-1}$, $\gamma_X=0.01$ ps$^{-1}$, the coupling strengths  
  $\hbar \Omega_R$ for panels (b), (c) and (d) were approximately equal to $0.98$ meV, $0.65$ meV and $0.32$ meV, respectively. }
\label{fig:fig_1}
\end{figure*}

Here, we experimentally confirm the existence of natural exceptional points in an exciton-polariton system by following a closed loop in parameter space, changing the external pumping power and particle momentum. We unequivocally demonstrate that exceptional points are separate from condensation and photon lasing critical points. We achieve this by using an  experimental technique where a strong resonant laser pump is used to excite up-converted excitations, which are observed directly in the transmission spectrum, as schematically shown in Fig.~\ref{fig:fig_1}(a). An analogous method was used to observe collective Bogoliubov excitations~\cite{Richard_stepanov2019dispersion,Pieczarka_2015, Pieczarka_2020, Biega_2021,Claude_2022}. In our case, photon lasing threshold is absent and the coherence is directly inherited from the pumping laser. In result, we are able to follow the path in parameter space without experiencing any jumps in the state of the system. We demonstrate analytically that in the limit of weak exciton-exciton interactions excitations inherit the properties of polariton dispersion, including exceptional points. Moreover, we exploit an incoherent reservoir of excitations to tune the light-matter coupling by more than one order of magnitude simply by changing the pump power. Thereby, we demonstrate an optically induced exceptional point in a generic exciton-polariton system without using additional degrees of freedom such as exciton spin, photon polarization, or a spatial degree of freedom. Our technique opens the way to investigation of a new class of models in polariton systems, where light-matter coupling can be manipulated locally by optical means. Possible applications of this technique include  investigations of dissipative phase transitions~\cite{Littlewood_PhysRevResearch.2.033018} and non-Hermitian topological states~\cite{Ueda_PhysRevX.8.031079,Comaron_PhysRevResearch.2.022051}.

\section{Results}

The idea of our experiment can be summarized in Fig.~\ref{fig:fig_1}(a). We investigate a semiconductor microcavity based on CdTe. In this structure photons confined in the optical resonator can strongly couple to excitons in quantum wells and form mixed states called exciton-polaritons. We drive the cavity with a continuous wave laser beam resonantly into the polariton mode in a transmission configuration. In addition to the transmitted laser, we observe an up-converted emission coming out of the sample. The emitted light follows exciton-polariton dispersion relation consisting of lower ($E_-$) and upper ($E_+$) polariton branches. When we increase the laser power ($P_2>P_1$), the observed polariton dispersion changes, corresponding to a decrease of the exciton-photon coupling strength. In this way, for sufficiently high excitation powers, we can suppress the coupling strength and drive the system into the weak coupling regime. At a specific laser power ($P_3$), matching exactly to the transition between the two coupling regimes, we can expect the appearance of exceptional points (EPs) in the dispersion relation. To explain the effects resulting from the presence of the EPs, we first analyze the system theoretically.

We first consider a planar semiconductor microcavity illustrated in Figs.~\ref{fig:fig_1}(a--c), with active medium sandwiched between two mirrors. Due to the spatial invariance in the plane of the cavity, the in-plane momentum of photons ${\bf k}=(k_x,k_y)$ is a constant of motion. To illustrate natural exceptional points it is sufficient to consider a simple, scalar toy model of light-matter coupling described with the non-Hermitian Hamiltonian
\begin{equation}
    \hat{H}= \begin{pmatrix}
E_C-i\hbar\gamma_C & \frac{\hbar}{2}\Omega_R \\
\frac{\hbar}{2}\Omega_R & E_X-i\hbar\gamma_X
\end{pmatrix},  
\label{eq:model}
\end{equation}
 where $E_C({\bf k})\approx E_C^0+(\hbar^2 {\bf k}^2/2 m_C)$ is the cavity photon energy dispersion with $m_C$ the effective mass of the photon, $E_X({\bf k})$ is the dispersion of the exciton which is approximately constant due to a large exciton effective mass, $\Omega_R$ is the light-matter (Rabi) coupling strength, while $\gamma_C$ and $\gamma_X$ are the photon and exciton decay rates, respectively~\cite{Yamamoto_RMP}.  Eigenmodes of the above Hamiltonian have complex energies $E_{\pm}= \frac{1}{2}\Big(E_C+E_X-i\hbar(\gamma_X+\gamma_C)\pm\sqrt{(\delta-i\hbar(\gamma_C-\gamma_X))^2+(\hbar\Omega_R)^2}\Big)$ where the photon-exciton detuning $\delta({\bf k}) = E_C({\bf k})-E_X({\bf k})$. Note that the effective exciton decay rate $\gamma_X$ may be negative if external pumping is stronger than the intrinsic exciton decay. 
 
 When the coupling strength is higher than the difference between decay rates of the exciton and photon modes, $\Omega_R >|\gamma_C-\gamma_X|$, the two eigenmodes are polariton modes with characteristic anticrossing in the $k$-dependent spectrum, as shown in Fig.~\ref{fig:fig_1}(d). Imaginary parts of energy or gain/loss rates of the two polariton modes continuously vary between the values corresponding to bare photon and exciton.
At the critical coupling strength $\Omega_R =|\gamma_C-\gamma_X|$, spectrum exhibits exceptional points as shown in~Fig.~\ref{fig:fig_1}(e). These are the only points where eigenvalues and eigenvectors coalesce as the matrix in Eq.~(\ref{eq:model}) becomes degenerate. They appear when the $k$-dependent detuning $\delta( {\bf k})$ is equal to zero. In an isotropic planar cavity, exceptional points form a ring in the $(k_x,k_y)$ space. Note that in a nonlinear system exceptional points may occur at positive detuning due to blueshift of exciton energy~\cite{Littlewood2019}. Finally, when the coupling strength  $\Omega_R <|\gamma_C-\gamma_X|$, spectrum is typical for weakly coupled system, as shown in Fig.~\ref{fig:fig_1}(f), where the two modes are exciton and photon modes with only a small admixture of the other. In this case, imaginary part of energy is almost independent of momentum and corresponds to exciton and photon decay rates. 

The exceptional point in the above described transition from weak to strong coupling coincides with the strong coupling criterion discussed in previous theoretical works~\cite{SAVONA1995733}. However, from the practical point of view, strong coupling can be also defined as the situation when broadening of spectral lines is smaller than mode splitting, so that upper and lower polariton modes are distinguishable, $\Omega_R \gtrsim \gamma_C, \gamma_X$. It is important to realize that these two criteria for strong coupling do not coincide in general, in particular when exciton and photon decay rates are comparable. 

The above described critical point is not the same as the threshold for photon lasing~\cite{Littlewood2019,Khurgin2020}. Indeed, the simple model given by~Eq.~(\ref{eq:model}) does not require a macroscopically occupied state and is valid also in the case of vacuum fluctuations. We demonstrate experimentally the occurrence of natural exceptional points in a system where there is {\it no transition to photon lasing at all}. We show that the exceptional points described above appear, under certain conditions, in the linearized spectrum of elementary excitations of a quantum fluid of light excited with an external resonant laser. In this setup, the macroscopically populated polariton state is present at any pumping power, as it is imposed by the external laser. Consequently, this system does not undergo a transition from a non-macroscopic to a macroscopic state in function of pumping. Moreover, the experimental technique that we use allows us to tune the Rabi coupling strength $\Omega_R$ in a wide range. This is achieved with the influence of a reservoir of incoherent particles created by the pump laser.


\begin{figure*}
\centering
  \includegraphics[width=\textwidth]{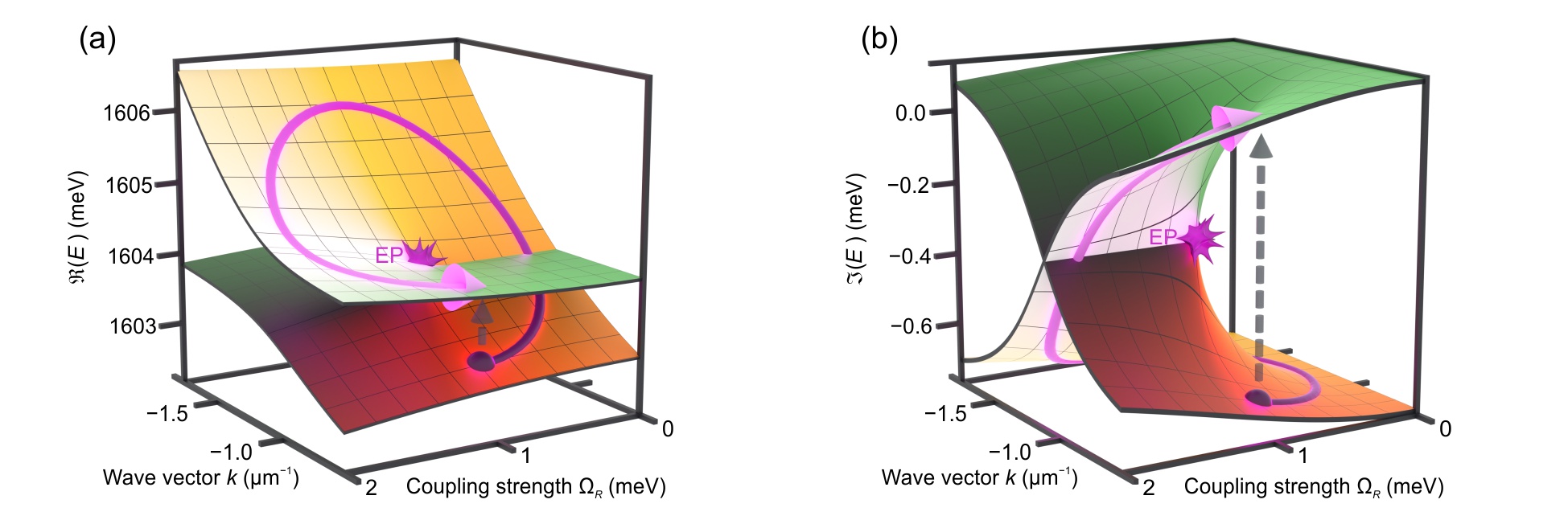}
  \caption{Encirclement of an exceptional point. Riemann surfaces showing (a) real and (b) imaginary part of the eigenenergy of Hamiltonian~(\ref{eq:model}) in function of  pump-dependent light-matter coupling strength $\Omega_R$ and  polariton in-plane momentum $k$. A single $2\pi$ encirclement of an exceptional point results in an exchange of eigenstates between the two complex branches of solutions.
  } \label{fig:fig_2}
\end{figure*}

We model the system with coupled equations for exciton and photon coherent fields described by $\psi_X$ and $\psi_C$ \begin{equation}
i\hbar\frac{d}{dt}\psi_C=(E_C({\bf k})-i\hbar\gamma_C)\psi_C+\frac{\hbar\tilde{\Omega}_R}{2}\psi_X+Fe^{-i\omega_pt},
\label{eq:Av.1}
\end{equation}
\begin{equation}
i\hbar\frac{d}{dt}\psi_X=(E_X({\bf k})-i\hbar\gamma_X)\psi_X+\frac{\hbar\tilde{\Omega}_R}{2}\psi_C+g_X|\psi_X|^2\psi_X,
\label{eq:Av.2}
\end{equation}
where  $F$ is the amplitude of the coherent laser field driving the cavity at frequency $\omega_p$. For simplicity, we neglect the effect of spatial inhomogeneity of the system and of the pump beam. The interactions between excitons are described with the coefficient $g_X$. 
Parameter $\tilde{\Omega}_R$ describes the coupling between excitons and photons including the effects of saturation~\cite{Schwendimann_2000}. Saturation may result from various effects, including the presence of a reservoir of incoherent excitons or free carriers. For example, Pauli exclusion results in phase-space filling leading to the reduction of exciton-photon coupling and the appearance of  anharmonic interaction between exciton and photon modes~\cite{Ciuti_2003}. Another possible source of saturation is the screening of free carriers. Moreover, it was demonstrated in several works that resonant pumping of a polariton system may result in the creation of a reservoir~\cite{Dominici2015,Hofling_Dark,Krizhanovskii_DarkSolitons}, despite particles in the  reservoir having higher energy than the pump laser.
Saturation may also result from the breakdown of bosonic description of polaritons in the limit of high density of particles.
Since exciton-polaritons are composite bosons formed as a pairs of fermions, 
the fermionic character of polaritons components manifests itself when the distance between excitons becomes comparable to their Bohr radius $a_B$~\cite{Schwendimann_2000}. 

We phenomenologically model saturation as a reduction of coupling strength,
$\tilde{\Omega}_R= \Omega_R^0e^{-(\beta_R n_R+\beta_X|\psi_X|^2)}$,
where $n_R$  is the reservoir density, while parameters $\beta_R$ and $\beta_X$ determine the strength of saturation. We assume that in the steady state $n_R$ depends linearly on the ratio of the optical pump intensity and the decay rate of the reservoir $n_R\approx\alpha\frac{|F|^2}{\gamma_R}$. We additionally assume that the reduction of coupling strength is mainly induced by the high energy incoherent excitons in the reservoir, which are not directly coupled to the photon field. We neglect the effects of stimulated scattering from the reservoir to the polariton fluid~\cite{Bobrovska_Adiabatic} as negligible compared to the resonant pumping rate. These assumptions allow us to rewrite the effective coupling strength in the first order approximation as
\begin{equation}\label{eq:4}
\tilde{\Omega}_R= \Omega_R^0(1-\beta n_R) = \Omega_R^0\left(1- \frac{\beta }{\gamma_R}|F|^2\right),
\end{equation}
where $\beta=\alpha\beta_R$.

Here, a large population of exciton-polaritons is created by a laser pump tuned to the energy of the lower polariton dispersion minimum at ${\bf k}=0$. It excites a  polariton quantum fluid with approximately zero in-plane momentum. 
By monitoring the transmission spectrum, in addition to the main pump laser, we observe  dispersion lines at nonzero momenta and energy higher than the lower polariton minimum, see Fig.~\ref{fig:fig_1}(a). This additional emission corresponds to fluctuations existing on top of the quantum fluid. These fluctuations result from up-conversion of polariton modes to excited states due to interactions. 

With the increase of pump power, we observe  reduction of splitting between lower and upper polariton modes while no signinificant blueshift is observed. This points out that, in contrast to previous experiments with nonresonant pumping, the effect of reduction of light-matter coupling is much stronger than the interaction driven blueshift. Under the approximation of negligible exciton-exciton interaction, we derive the effective non-Hermitian Hamiltonian for the main branch of linearized excitations (see supplementary information)
 \begin{equation}
    \hat{H}= \begin{pmatrix}
E_C(k)-\hbar \omega_p-i\hbar\gamma_C & \frac{\hbar}{2}\tilde{\Omega}_R \\
\frac{\hbar}{2}\tilde{\Omega}_R & E_X-\hbar \omega_p-i\hbar\gamma_X
\end{pmatrix}.  \label{eq:Bogo}
\end{equation}
The above model resembles the Hamiltonian of the original polariton model, Eq.~(\ref{eq:model}). Therefore, Bogoliubov excitations inherit the properties of vacuum polariton modes, including the transition through exceptional points. They can be used as a probe of the intrinsic properties of the system, even when the main transmission signal is dominated by the pump laser.

We now demonstrate the experimental setup that can be used to perform encirclement of an exceptional point in the parameter space. Previously, it was suggested that this can be achieved by tuning the detuning and pump power in parameter space along a closed path~\cite{Littlewood2019,Khurgin2020}. However, in the case of nonresonantly pumped condensates, this is very challenging as the emission is dominated by the strong emission from the bosonic condensate, and the reduction of coupling strength can be achieved only at very high density, close to the Mott transition density, where the spectrum is dramatically modified by other effects. In contrast, with our resonant pumping technique we are able to tune the coupling strength smoothly and widely across the transition from weak to strong coupling while monitoring the changes of the excitation spectrum. 

In principle, encirclemnt of an exceptional point or exceptional ring requires tuning two parameters, the detuning $\delta$ and the coupling strength $\Omega_R$. We exploit the direct dependence between $\delta({\bf k})$ and the polariton momentum ${\bf k}$ to observe encirclement by tuning only a single parameter, the coupling strength $\Omega_R$. This is illustrated in Fig.~\ref{fig:fig_2}, which shows the depndence of (a)~real and (b)~imaginary part of the energy in Eq.~(\ref{eq:model}) in the vicinity of one of the exceptional points as a function of the transverse momentum and coupling strength. A clear point of degeneracy is visible. By following a closed loop around this exceptional point in the ($k$,$\Omega_R$) plane, performing a $2\pi$ encirclement, one ends up at a different branch of the non-Hermitian, complex spectrum. Another full encirclement is required to return to the original state. This exchange of states after a full encirclement is a characteristic property of exceptional points and has no analog in Hermitian physics.

One of the parameters that are required for performing the loop, the momentum $k$, does not have to be externally tuned during the experiment. This results from the experimental technique that allows to detect energies for all $k$ vectors at the same time. Therefore, to record all energies necessary to perform a full loop in the ($k$,$\Omega_R$) parameter space, we need to tune $\Omega_R$ only. 


\begin{figure*}
\centering
\includegraphics[width=\textwidth]{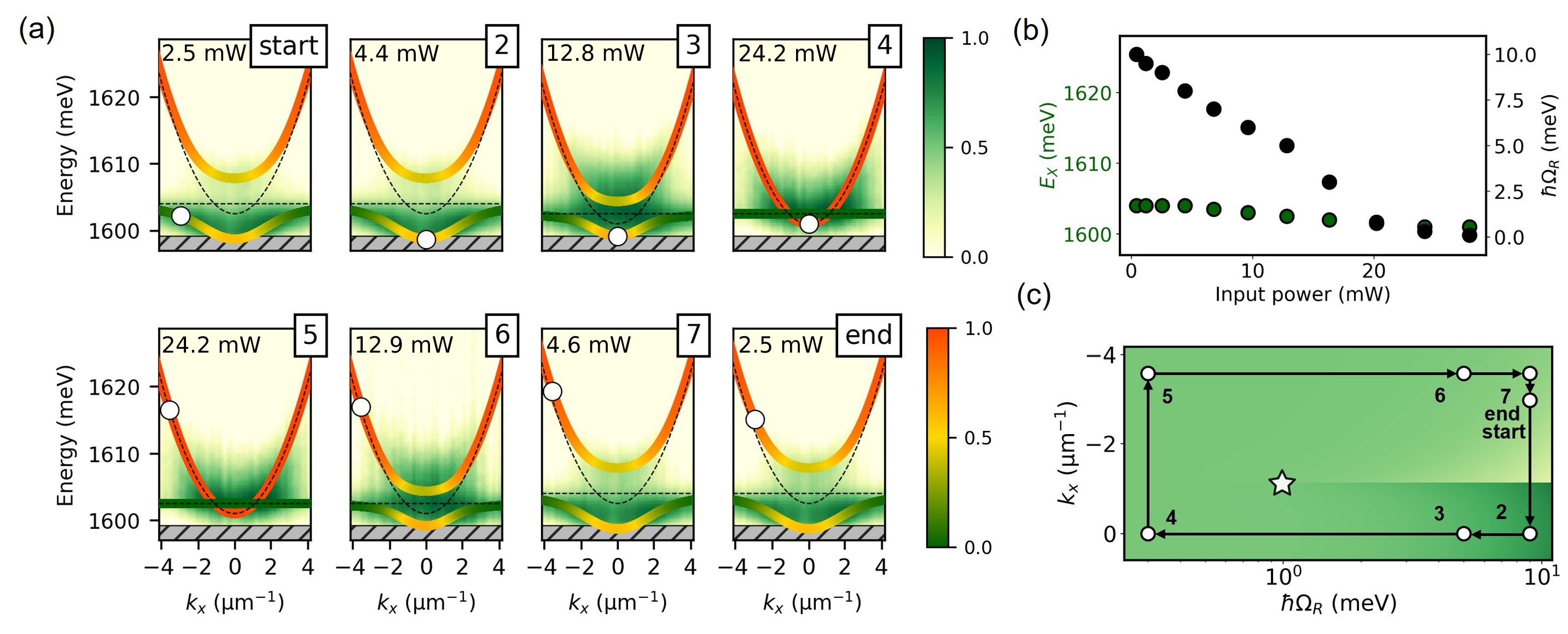}
  \caption{Experimental spectra. (a) Transmission spectra of quantum fluid of light excited with a resonant pump at in-plane momentum ${\bf k}=0$. The spectra show results of experiments at different pumping intensities (changing the power in the sequence: "start"$\rightarrow$2.5\,mW, "2"$\rightarrow$4.4\,mW, "3"$\rightarrow$12.8\,mW, "4" and "5"$\rightarrow$24.2\,mW, "6"$\rightarrow$12.9\,mW, "7"$\rightarrow$4.6\, mW, "end"$\rightarrow$2.5\,mW) corresponding to different coupling strengths $\Omega_R$. Coloured thick lines are the fitted energies of the modes according to Eq.~(\ref{eq:Bogo}) and the color scale of the lines corresponds to the exciton/photon composition of the eigenmode. (b) Coupling strength and exciton energy as a function of laser power incident resonantly at the lower polariton mode. The values were obtained by fitting the exciton-polaritons dispersion to the angle-resolved maps. (c)  Phase diagram and the closed loop with points marked by numbers corresponding to white circles in (a). White star marks the point in parameter space where exceptional points occur.}
\label{fig:fig_3}
\end{figure*}

\begin{figure}
\centering
  \includegraphics[width=7.5cm,height=11.0cm]{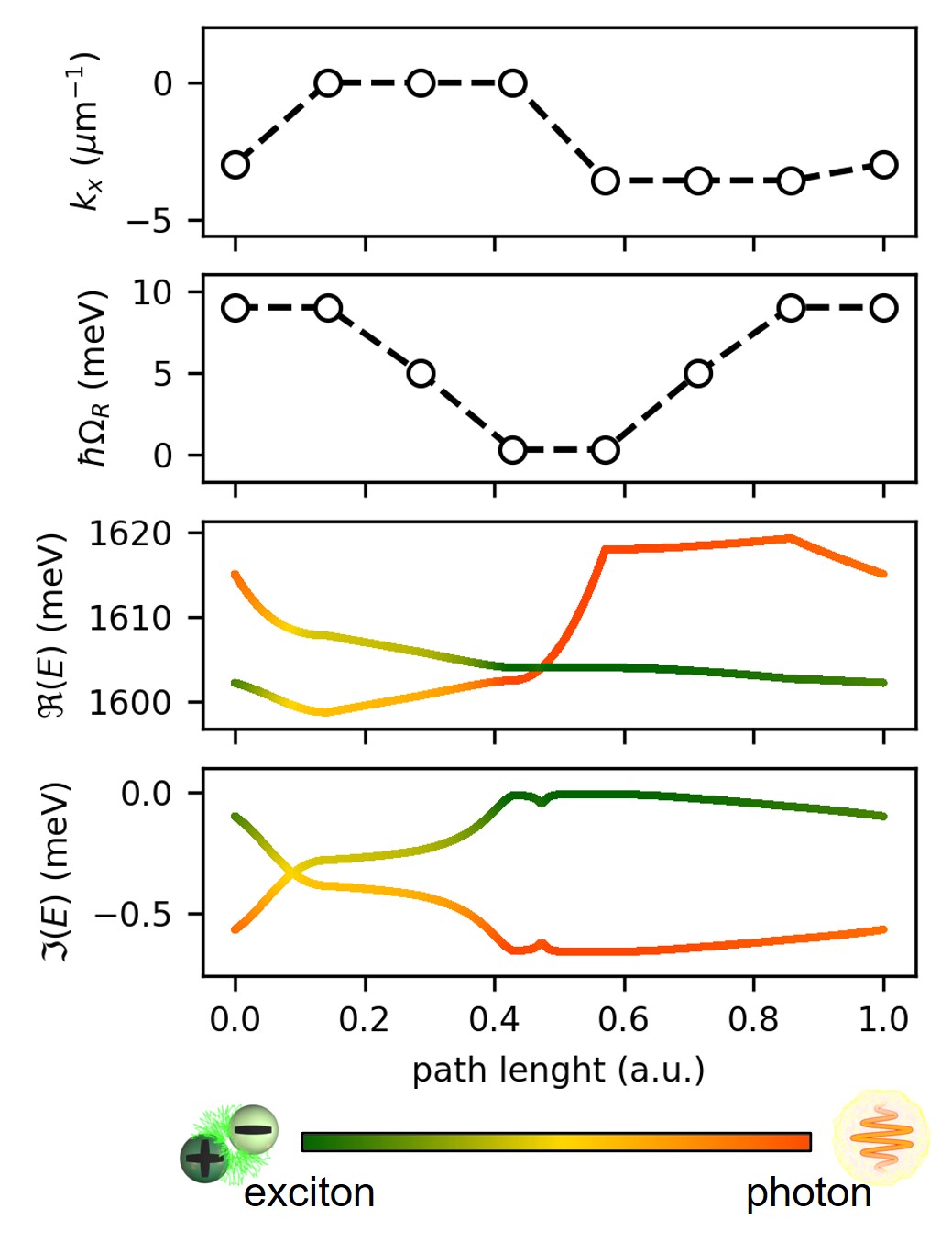}
  \caption{ Real and imaginary parts of eigenenergies. (From top to bottom) The calculated momentum, coupling strength, and real and imaginary parts of eigenenergies along the closed path covered in the experiment. The colour scale shows the composition of the eigenmodes, where red and green correspond to the pure photon and exciton modes.}
\label{fig:fig_4}
\end{figure}


Experimental spectra are presented in Fig.~\ref{fig:fig_3}(a). Dashed lines show fits to the theoretical model of Eq.~(\ref{eq:Bogo}). The corresponding values of Rabi coupling and exciton energy are shown in~Fig.~\ref{fig:fig_3}(b). We are able to tune the coupling strength by more than one order of magnitude simply by varying the intensity of the pump laser. 
We attribute this reduction of coupling strength to a resonance of the pump laser with a biexciton mode that results in a fast buildup of an incoherent reservoir~\cite{Deveaud_Bistability}. In our structure, we expect that biexciton energy is located between 3.3--4.6\,meV below the free exciton~\cite{Adachi_PhysE2001}.

White circles in Fig.~\ref{fig:fig_3}(a) denote the points on the closed loop in the ($k$,$\Omega_R$) parameter space corresponding to the points in the phase diagram in Fig.~\ref{fig:fig_3}(c). The loop encircles the exceptional point marked with a star. The path follows the eigenmodes of the system continuously, as each point can be reached from the previous one by a continuous change of $k$ or pump power with no discontinuity in between. It first follows the lower polariton branch at fixed pump power (start$\rightarrow$2), then stays at the minimum of dispersion as the pump power is increased, crossing into the weak coupling regime (2$\rightarrow$3$\rightarrow$4), follows the photonic branch to negative $k$ (4$\rightarrow$5) and stays at the photonic branch as the pump power is increased while the system returns to the strong coupling regime (5$\rightarrow$6$\rightarrow$7). The end point of the path corresponds to the same parameters ($k$,$\Omega_R$) as the starting point, but now the point lies on the upper polariton branch. Clearly, we can perform another full encirclement of the exceptional point to return to the original point on the lower polariton branch. Fig.~\ref{fig:fig_4} shows the  calculated $k$, $\Omega_R$, and real and imaginary parts of the energy along the closed loop. A clear exchange of eigenstates is also visible, manifesting itself in the exchange of both real and imaginary parts of the complex eigenenergies.

\section{Conclusion}

In conclusion, we demonstrated that perturbative spectra of excitations of a resonantly pumped condensate exhibit natural exceptional points occurring at the transition from weak to strong coupling regime of light-matter interaction. The presented experiment clearly demonstrates that this transition does not, in general, coincide with condensation to photon lasing transition, confirming previous theoretical results. Our results demonstrate the potential of this experimental technique not only to probe the properties of quantum fluids of light, but also for tuning coupling strength by more that one order of magnitude. The possibility to independently tune light-matter coupling strength and the detuning between exciton and photon modes allowed to perform an encirclement of an exceptional point. Importantly, this tunability is independent of the blueshift and can be controlled locally by an appropriately shaped pump beam. The ability to locally control light-matter coupling with an independent knob opens the way to implementation of a new class of non-Hermitian Hamiltonians, exhibiting anomaluos topological states~\cite{Ueda_PhysRevX.8.031079} or dissipative phase transitions~\cite{Littlewood_PhysRevResearch.2.033018}.

\section{Methods}

\subsection{Details of the microcavity sample} 

\noindent The experiment used a semiconductor microcavity based on elements of groups II-VI of the periodic table. The cavity was grown using Molecular Beam Epitaxy and consists of two Bragg mirrors. Each mirror contains 20 pairs of alternately arranged CdTe layers alloyed with zinc and magnesium with various ratios. The thickness of the cavity between the mirrors is around $4\lambda/2n = 600$~nm. 6 CdTe:Mn quantum wells were placed at the maxima of electric field distribution within the microcavity, ensuring a strong coupling regime.
The cavity was grown on the nontransparent GaAs epitaxial substrate, with an additional 90~nm thick MgTe "sacrificial layer" between the cavity and the substrate. Following the water exfoliation procedure~\cite{Seredynski_2018}, the transmissive microcavity structure was transferred from the initial substrate to transparent Al\textsubscript{2}O\textsubscript{3}. The typical coupling strength for this sample at 5~K, determined based on angle-resolved reflectance at low power, was around $\Omega_R = 10$~meV. The \mbox{Q-factor} of the structure was equal to 345~\cite{Furman_2023}. The linewidths of the bare cavity mode and of the exciton resonance were equal to 6.46\,meV and 10.74\,meV, correspondingly.

\subsection{Optical experiment}
\noindent Measurements were conducted at a temperature of 5~K with the microcavity sample placed inside a cryostat. The sample was excited in transmission configuration with a tunable, continuous wave, Ti:sapphire laser with wavelength of $\lambda=776$~nm (1.598~meV), tuned to the bottom of the lower polariton mode.
The laser beam was focused with a long focal length lens ($f = 25.4$~mm) to ensure the excitation at zero wave vector. The laser spot on the sample had a diameter of $d=12~\mu$m. To collect the transmitted light we used a microscopic objective with a high numerical aperture ($\text{NA}=0.55$). Two convex lenses were used to image the Fourier plane of the microscope objective on the entrance slit of a spectrometer, to collect spectra resolved in photon momenta and energy. 
In contrast to a pump-probe experiment --- where two different laser pulses are used, one to excite a sample, and the other to probe the resulting changes --- in our experiment we used only one laser to induce emission from the structure. The light coming out from the sample was passed through a longpass filter 
to cut out the scattered laser and to collect only the up-converted  spectrum at higher photon energies and nonzero momenta. The spectral range blocked by the filter is presented by the hatched area in Fig.~\ref{fig:fig_3}(a). The measured  spectra for varying laser beam pumping powers are shown in Fig.~\ref{fig:fig_3}(a). Map 1, marked "start", presents the result acquired at a low pumping power (2.5\,mW), when the system is in strong coupling regime and the observed coupling strength is equal to $\Omega_R = 9.8$~meV. Maps 2$\rightarrow$4 show the dispersion relations for increasing pumping power (4.4\,mW,  12.8\,mW and 24.2\,mW correspondingly), which results in decreasing exciton-photon coupling strength. This finally leads to the loss of strong coupling in panel 4. Subsequently, when the pumping power is gradually reduced (5 to "end"), the coupling strength increases, and the strong coupling regime is finally recovered (map "end").

\begin{backmatter}
\bmsection{Funding}
\noindent This work is supported by National Science Center, Poland:\\ A.O. acknowledges 2019/35/N/ST3/01379, R.M. acknowledges 2019/33/N/ST3/02019, M.K. acknowledges 2018/31/N/ST3/03046, M.M. acknowledges 2016/22/E/ST3/00045, B.P., M.F., K.T., J. S. acknowledges 2020/37/B/ST3/01657.

\bmsection{Disclosures}
\noindent The authors declare no conflicts of interest.

\bmsection{Data availability}
\noindent Data underlying the results presented in this paper could be obtained from the authors upon reasonable request.
\bmsection{Supplemental document} See Supplementary information for supporting content.

\appendix
\section*{Supplementary information}

\section{Analysis of linear excitations}
The effective non-Hermitian Hamiltonian described in the main text can be obtained using the approximation of linearized excitations, considering small fluctuations of the photonic $\delta_C$ and excitonic $\delta_X$ fields around their steady states  $\psi_C^{ss}$ and $\psi_X^{ss}$. The derivation goes is analogous to the derivation of Bogoliubov-de Gennes modes in an interacting Bose gas. In our analysis, the mean-field photonic and excitonic polariton components were 
\begin{eqnarray}
   \psi_C({\bf{r}},t)&=\psi_C^{ss}e^{-i\omega_pt}(1+\delta_C({\bf{r}},t)),
   \label{eq:Ap.1}\\
   \psi_X({\bf{r}},t)&=\psi_X^{ss}e^{-i\omega_pt}(1+\delta_X({\bf{r}},t)).
   \label{eq:Ap.2}
\end{eqnarray}

Substituting Eqs.~\eqref{eq:Ap.1} and \eqref{eq:Ap.2} into the system of Eqs.~(2) and (3), the evolution of particle-like $\delta_X$ and $\delta_C$ and hole-like excitations, $\delta_X^*$ and $\delta_C^*$ were derived
\begin{eqnarray}
    i\hbar\frac{\partial}{\partial t}\delta_C&=(\hat{E}_C-\hbar\omega_p-i\hbar\gamma_C)\delta_C+\frac{\hbar \tilde{\Omega}_R}{2}\frac{\psi_X^{ss}}{\psi_C^{ss}}\delta_X,
    \label{eq:Ap.3}\\
    i\hbar\frac{\partial}{\partial t}\delta^*_C&=-(\hat{E}_C-\hbar\omega_p+i\hbar\gamma_C)\delta^*_C-\frac{\hbar \tilde{\Omega}_R}{2}\frac{\psi_X^{ss}}{\psi_C^{ss}}\delta^*_X,
    \label{eq:Ap.4}
\end{eqnarray}
\begin{eqnarray}
    i\hbar\frac{\partial}{\partial t}\delta_X&=(\hat{E}_X-\hbar\omega_p-i\hbar\gamma_X)\delta_X+\frac{\hbar \tilde{\Omega}_R}{2}\frac{\psi_C^{ss}}{\psi_X^{ss}}\delta_C+g_X(\psi_X^{ss})^2(2\delta_X+\delta_X^*),
    \label{eq:Ap.5}\\
    i\hbar\frac{\partial}{\partial t}\delta^*_X&=-(\hat{E}_X-\hbar\omega_p+i\hbar\gamma_X)\delta^*_X-\frac{\hbar \tilde{\Omega}_R}{2}\frac{\psi_C^{ss}}{\psi_X^{ss}}\delta^*_C-g_X(\psi_X^{ss})^2(2\delta^*_X+\delta_X).
    \label{eq:Ap.6}
\end{eqnarray}

Keeping only the linear terms, Eqs.~\eqref{eq:Ap.3}--\eqref{eq:Ap.6} can be written in a matrix form as
\begin{equation}
\scalemath{0.7}{i\hbar\frac{\partial}{\partial t}\begin{pmatrix}
\delta_C(t,k)\\
\delta_X(t,k) \\
\delta_C^{*}(t,k)\\
\delta_X^{*} (t,k)\\
\end{pmatrix}=
\begin{pmatrix}
E_C(k)-\hbar\omega_p-i\hbar\gamma_C & \frac{\hbar \tilde{\Omega}_R}{2}\frac{\psi_X^{ss}}{\psi_C^{ss}} & 0                  & 0 \\\frac{\hbar \tilde{\Omega}_R}{2}\frac{\psi_C^{ss}}{\psi_X^{ss}} &E_X(k)-\hbar\omega_p-i\hbar\gamma_X +2g_X(\psi_X^{ss})^2 &0&g_X(\psi_X^{ss})^2\\
0 & 0 & -E_C(k)+\hbar\omega_p-i\hbar\gamma_C &-\frac{\hbar \tilde{\Omega}_R}{2}\frac{\psi_X^{ss}{}^{*}}{\psi_C^{ss}{}^{*}} \\
0 & -g_X(\psi_X^{ss})^2 & -\frac{\hbar \tilde{\Omega}_R}{2}\frac{\psi_C^{ss}{}^{*}}{\psi_X^{ss}{}^{*}} &-E_X(k)+\hbar\omega_p-i\hbar\gamma_X -2g_X(\psi_X^{ss})^2 \\
\end{pmatrix}\begin{pmatrix}
\delta_C(t,k)\\
\delta_X(t,k) \\
\delta_C^{*}(t,k)\\
\delta_X^{*} (t,k)\\
\end{pmatrix}.
}
\label{eq:Ap.7}
\end{equation}
Eigenenergies of the matrix are equivalent to the spectrum of elementary excitations of the system. Assuming that the interaction strength between excitons is negligibly small  ($g_X\approx0$), considering the part of the spectrum near the exceptional point where $\frac{\psi_C^{ss}}{\psi_X^{ss}}\approx\frac{\psi_X^{ss}}{\psi_C^{ss}}\approx 1$, $\frac{\psi_C^{ss}{}^*}{\psi_X^{ss}*}\approx\frac{\psi_X^{ss}{}^*}{\psi_C^{ss}{}^*} \approx 1$, Eq.~(\ref{eq:Ap.7}) takes the form
\begin{equation}
\scalemath{0.8}{i\hbar\frac{\partial}{\partial t}\begin{pmatrix}
\delta_C(t,k)\\
\delta_X(t,k) \\
\delta_C^*(t,k)\\
\delta_X^* (t,k)\\
\end{pmatrix}=
\begin{pmatrix}
E_C(k)-\hbar\omega_p-i\hbar\gamma_C & \frac{\hbar \tilde{\Omega}_R}{2}& 0                  & 0 \\\frac{\hbar \tilde{\Omega}_R}{2} &E_X(k)-\hbar\omega_p-i\hbar\gamma_X  &0&0\\
0 & 0 & -E_C(k)+\hbar\omega_p-i\hbar\gamma_C &-\frac{\hbar \tilde{\Omega}_R}{2} \\
0 & 0 & -\frac{\hbar \tilde{\Omega}_R}{2} &-E_X(k)+\hbar\omega_p-i\hbar\gamma_X  \\
\end{pmatrix}\begin{pmatrix}
\delta_C(t,k)\\
\delta_X(t,k) \\
\delta_C^*(t,k)\\
\delta_X^* (t,k)\\
\end{pmatrix},
}
\label{eq:Ap.8}
\end{equation}
The block structure of the above matrix reflects the ``normal'' and ``ghost'' branches of the Bogoliubov excitation spectrum of elementary excitations in the system. Focusing on the normal branch, for $\delta_C$ and $\delta_X$ we recover the spectrum of the non-Hermitian Hamiltonian described by Eq.~(5).

\section{Collective excitation spectrum}
Bogoliubov quasiparticles are collective excitations emerging in superconductors and superfluids. Their signatures in bosonic condensates include the characteristic linear dependence between the energy of elementary excitation and the wave vector \cite{pitaevskii2003bose}. Additionally, Bogoliubov excitation spectrum is symmetric with respect to vacuum energy, which results in the appearance of a virtual energy branch, the so-called "ghost branch". 
 
According to the Landau criterion, a system with such a spectrum behaves like a superfluid. Therefore, observation of Bogoliubov excitation was essential for understanding the nature of bosons scattering, interactions, and their condensates stability.

In the case of nonequilibrium, incoherently pumped exciton-polariton condensates, Bogoliubov excitation has a more complex character than in the equilibrium case. The nontrivial shape of the spectra is induced by the interaction with an excitonic reservoir. Such an interaction can be observed even in resonantly excited systems~
\cite{Gavrilov_2010, Vishnevsky_2012, Sekretenko_2013,Richard_stepanov2019dispersion}. During the past several years, these interesting observations have stimulated numerous questions about the nature of collective excitations in polariton condensates.

According to the theoretical framework presented in a recent work~\cite{Richard_stepanov2019dispersion}, these effects can result from two-body interactions. This work experimentally confirmed that the excitation observed in such a system exhibits a hybrid nature connecting properties of the Bogoliubov and reservoir density excitations. The appearance of an additional excitonic reservoir can drastically affect the superfluid properties and change the character of the quantum fluid spectrum.

Here, we want to point out some similarities between our experimental observation and the results of the previous work\cite{Richard_stepanov2019dispersion}. Similarly as in the previous work~\cite{Richard_stepanov2019dispersion}, we assume that thermally excited acoustic phonons generate an incoherent reservoir which interacts with polaritons. 
From the experimental point of view, the main difference between our results and the previous work is in the chosen material of the quantum well. Compared to the GaAs structure explored in the previous work, our CdTe-based microcavity is characterized by the lack of observable energy blueshift of polariton modes. 

In result, the excitation spectrum in our case is substantially modified as the laser power is increased, with a visible reduction of coupling strength and polariton up-conversion. However, due to the weakness of interactions, the effects characteristic for the Bogoliubov quasiparticles, such as the linearized energy spectrum at low momenta and the ghost branch, cannot be detected.

It should be noted that the exact physical mechanism for the observed up-conversion, and the role of biexciton or dark exciton reservoir are under debate and will stimulate further discussion on this subject.

\end{backmatter}

\bibliography{bib}

\begin{thebibliography}{10}
\newcommand{\enquote}[1]{``#1''}

\bibitem{Littlewood2019}
R.~Hanai, A.~Edelman, Y.~Ohashi, and P.~B. Littlewood, \enquote{{Non-Hermitian
  Phase Transition from a Polariton Bose-Einstein Condensate to a Photon
  Laser},} {\protect\JournalTitle{Phys. Rev. Lett.}} \textbf{122}, 185301
  (2019).

\bibitem{Khurgin2020}
J.~B. Khurgin, \enquote{{Exceptional points in polaritonic cavities and
  subthreshold Fabry--Perot lasers},} {\protect\JournalTitle{Optica}}
  \textbf{7}, 1015--1023 (2020).

\bibitem{breuer2016colloquium}
H.-P. Breuer, E.-M. Laine, J.~Piilo, and B.~Vacchini, \enquote{Colloquium:
  {Non-Markovian} dynamics in open quantum systems,}
  {\protect\JournalTitle{Reviews of Modern Physics}} \textbf{88}, 021002
  (2016).

\bibitem{rivas2012open}
A.~Rivas and S.~F. Huelga, \emph{Open quantum systems}, vol.~10 (Springer,
  2012).

\bibitem{Devoret_PhysRevA.88.023849}
Z.~Leghtas, U.~Vool, S.~Shankar, M.~Hatridge, S.~M. Girvin, M.~H. Devoret, and
  M.~Mirrahimi, \enquote{{Stabilizing a Bell state of two superconducting
  qubits by dissipation engineering},} {\protect\JournalTitle{Phys. Rev. A}}
  \textbf{88}, 023849 (2013).

\bibitem{Verstraete2009}
F.~Verstraete, M.~M. Wolf, and J.~Ignacio~Cirac, \enquote{Quantum computation
  and quantum-state engineering driven by dissipation,}
  {\protect\JournalTitle{Nature Physics}} \textbf{5}, 633--636 (2009).

\bibitem{bender2007making}
C.~M. Bender, \enquote{{Making sense of non-Hermitian Hamiltonians},}
  {\protect\JournalTitle{Reports on Progress in Physics}} \textbf{70}, 947
  (2007).

\bibitem{Heiss2012}
W.~D. Heiss, \enquote{The physics of exceptional points,}
  {\protect\JournalTitle{Journal of Physics A: Mathematical and Theoretical}}
  \textbf{45}, 444016 (2012).

\bibitem{Mohammad2019}
M.-A. Miri and A.~Al\`u, \enquote{Exceptional points in optics and photonics,}
  {\protect\JournalTitle{Science}} \textbf{363}, eaar7709 (2019).

\bibitem{Weitz2021}
F.~E. \"Ozt\"urk, T.~Lappe, G.~Hellmann, J.~Schmitt, J.~Klaers, F.~Vewinger,
  J.~Kroha, and M.~Weitz, \enquote{{Observation of a non-Hermitian phase
  transition in an optical quantum gas},} {\protect\JournalTitle{Science}}
  \textbf{372}, 88--91 (2021).

\bibitem{Szameit2015}
J.~M. Zeuner, M.~C. Rechtsman, Y.~Plotnik, Y.~Lumer, S.~Nolte, M.~S. Rudner,
  M.~Segev, and A.~Szameit, \enquote{{Observation of a Topological Transition
  in the Bulk of a Non-Hermitian System},} {\protect\JournalTitle{Phys. Rev.
  Lett.}} \textbf{115}, 040402 (2015).

\bibitem{Ueda_PhysRevX.8.031079}
Z.~Gong, Y.~Ashida, K.~Kawabata, K.~Takasan, S.~Higashikawa, and M.~Ueda,
  \enquote{{Topological Phases of Non-Hermitian Systems},}
  {\protect\JournalTitle{Phys. Rev. X}} \textbf{8}, 031079 (2018).

\bibitem{Hodaei2017}
H.~Hodaei, A.~U. Hassan, S.~Wittek, H.~Garcia-Gracia, R.~El-Ganainy, D.~N.
  Christodoulides, and M.~Khajavikhan, \enquote{Enhanced sensitivity at
  higher-order exceptional points,} {\protect\JournalTitle{Nature}}
  \textbf{548}, 187--191 (2017).

\bibitem{Wiersig2020}
J.~Wiersig, \enquote{Review of exceptional point-based sensors,}
  {\protect\JournalTitle{Photon. Res.}} \textbf{8}, 1457--1467 (2020).

\bibitem{Christodoulides_PhysRevLett.103.093902}
A.~Guo, G.~J. Salamo, D.~Duchesne, R.~Morandotti, M.~Volatier-Ravat, V.~Aimez,
  G.~A. Siviloglou, and D.~N. Christodoulides, \enquote{Observation of
  $\mathcal{P}\mathcal{T}$-symmetry breaking in complex optical potentials,}
  {\protect\JournalTitle{Phys. Rev. Lett.}} \textbf{103}, 093902 (2009).

\bibitem{Doiron}
C.~F. Doiron and G.~V. Naik, \enquote{Non-{H}ermitian selective thermal
  emitters using metal--semiconductor hybrid resonators,}
  {\protect\JournalTitle{Advanced Materials}} \textbf{31}, 1904154 (2019).

\bibitem{Feng2017}
L.~Feng, R.~El-Ganainy, and L.~Ge, \enquote{{Non-Hermitian photonics based on
  parity--time symmetry},} {\protect\JournalTitle{Nature Photonics}}
  \textbf{11}, 752--762 (2017).

\bibitem{Chen2017}
W.~Chen, {\c{S}}.~K. {\"O}zdemir, G.~Zhao, J.~Wiersig, and L.~Yang,
  \enquote{Exceptional points enhance sensing in an optical microcavity,}
  {\protect\JournalTitle{Nature}} \textbf{548}, 192--196 (2017).

\bibitem{Yang2014}
B.~Peng, {\c{S}}.~K. \"{O}zdemir, S.~Rotter, H.~Yilmaz, M.~Liertzer, F.~Monifi,
  C.~M. Bender, F.~Nori, and L.~Yang, \enquote{Loss-induced suppression and
  revival of lasing,} {\protect\JournalTitle{Science}} \textbf{346}, 328--332
  (2014).

\bibitem{Gao2021}
Y.~Li, X.~Ma, Z.~Hatzopoulos, P.~G. Savvidis, S.~Schumacher, and T.~Gao,
  \enquote{Switching off a microcavity polariton condensate near the
  exceptional point,} {\protect\JournalTitle{ACS Photonics}} \textbf{9},
  2079--2086 (2022).

\bibitem{Gao2015}
T.~Gao, E.~Estrecho, K.~Y. Bliokh, T.~C.~H. Liew, M.~D. Fraser, S.~Brodbeck,
  M.~Kamp, C.~Schneider, S.~H{\"o}fling, Y.~Yamamoto, F.~Nori, Y.~S. Kivshar,
  A.~G. Truscott, R.~G. Dall, and E.~A. Ostrovskaya, \enquote{{Observation of
  non-Hermitian degeneracies in a chaotic exciton-polariton billiard},}
  {\protect\JournalTitle{Nature}} \textbf{526}, 554--558 (2015).

\bibitem{Gao2018}
T.~Gao, G.~Li, E.~Estrecho, T.~C.~H. Liew, D.~Comber-Todd, A.~Nalitov,
  M.~Steger, K.~West, L.~Pfeiffer, D.~W. Snoke, A.~V. Kavokin, A.~G. Truscott,
  and E.~A. Ostrovskaya, \enquote{{Chiral Modes at Exceptional Points in
  Exciton-Polariton Quantum Fluids},} {\protect\JournalTitle{Phys. Rev. Lett.}}
  \textbf{120}, 065301 (2018).

\bibitem{Zhang2017}
D.~Zhang, X.-Q. Luo, Y.-P. Wang, T.-F. Li, and J.~Q. You, \enquote{Observation
  of the exceptional point in cavity magnon-polaritons,}
  {\protect\JournalTitle{Nature Communications}} \textbf{8}, 1368 (2017).

\bibitem{Kono_gao2018continuous}
W.~Gao, X.~Li, M.~Bamba, and J.~Kono, \enquote{Continuous transition between
  weak and ultrastrong coupling through exceptional points in carbon nanotube
  microcavity exciton--polaritons,} {\protect\JournalTitle{Nature Photonics}}
  \textbf{12}, 362--367 (2018).

\bibitem{Krol_AnnihilationEP}
M.~Kr{\'{o}}l, I.~Septembre, P.~Oliwa, M.~K{\k{e}}dziora, K.~{\L}empicka-Mirek,
  M.~Muszy{\'{n}}ski, R.~Mazur, P.~Morawiak, W.~Piecek, P.~Kula,
  W.~Bardyszewski, P.~G. Lagoudakis, D.~D. Solnyshkov, G.~Malpuech,
  B.~Pi{\k{e}}tka, and J.~Szczytko, \enquote{Annihilation of exceptional points
  from different {D}irac valleys in a 2{D} photonic system,}
  {\protect\JournalTitle{Nat. Commun.}} \textbf{13}, 5340 (2022).

\bibitem{Liao_EP}
Q.~Liao, C.~Leblanc, J.~Ren, F.~Li, Y.~Li, D.~Solnyshkov, G.~Malpuech, J.~Yao,
  and H.~Fu, \enquote{Experimental measurement of the divergent quantum metric
  of an exceptional point,} {\protect\JournalTitle{Phys. Rev. Lett.}}
  \textbf{127}, 107402 (2021).

\bibitem{su2021direct}
R.~Su, E.~Estrecho, D.~Biega{\'n}ska, Y.~Huang, M.~Wurdack, M.~Pieczarka, A.~G.
  Truscott, T.~C. Liew, E.~A. Ostrovskaya, and Q.~Xiong, \enquote{Direct
  measurement of a non-{H}ermitian topological invariant in a hybrid
  light-matter system,} {\protect\JournalTitle{Sci. Adv.}} \textbf{7}, eabj8905
  (2021).

\bibitem{Richard_stepanov2019dispersion}
P.~Stepanov, I.~Amelio, J.-G. Rousset, J.~Bloch, A.~Lema{\^\i}tre, A.~Amo,
  A.~Minguzzi, I.~Carusotto, and M.~Richard, \enquote{Dispersion relation of
  the collective excitations in a resonantly driven polariton fluid,}
  {\protect\JournalTitle{Nat. Commun.}} \textbf{10}, 3869 (2019).

\bibitem{Pieczarka_2015}
M.~Pieczarka, M.~Syperek, L.~Dusanowski, J.~Misiewicz, F.~Langer, A.~Forchel,
  M.~Kamp, C.~Schneider, S.~H\"ofling, A.~Kavokin, and
  G.~S\ifmmode~\mbox{\k{e}}\else \k{e}\fi{}k, \enquote{Ghost branch
  photoluminescence from a polariton fluid under nonresonant excitation,}
  {\protect\JournalTitle{Phys. Rev. Lett.}} \textbf{115}, 186401 (2015).

\bibitem{Pieczarka_2020}
M.~Pieczarka, E.~Estrecho, M.~Boozarjmehr, O.~Bleu, M.~Steger, K.~West, L.~N.
  Pfeiffer, D.~W. Snoke, J.~Levinsen, M.~M. Parish, A.~G. Truscott, and E.~A.
  Ostrovskaya, \enquote{Observation of quantum depletion in a non-equilibrium
  exciton--polariton condensate,} {\protect\JournalTitle{Nature
  Communications}} \textbf{11}, 429 (2020).

\bibitem{Biega_2021}
D.~Biega\ifmmode~\acute{n}\else \'{n}\fi{}ska, M.~Pieczarka, E.~Estrecho,
  M.~Steger, D.~W. Snoke, K.~West, L.~N. Pfeiffer, M.~Syperek, A.~G. Truscott,
  and E.~A. Ostrovskaya, \enquote{Collective excitations of exciton-polariton
  condensates in a synthetic gauge field,} {\protect\JournalTitle{Phys. Rev.
  Lett.}} \textbf{127}, 185301 (2021).

\bibitem{Claude_2022}
F.~Claude, M.~J. Jacquet, R.~Usciati, I.~Carusotto, E.~Giacobino, A.~Bramati,
  and Q.~Glorieux, \enquote{High-resolution coherent probe spectroscopy of a
  polariton quantum fluid,} {\protect\JournalTitle{Phys. Rev. Lett.}}
  \textbf{129}, 103601 (2022).

\bibitem{Littlewood_PhysRevResearch.2.033018}
R.~Hanai and P.~B. Littlewood, \enquote{Critical fluctuations at a many-body
  exceptional point,} {\protect\JournalTitle{Phys. Rev. Research}} \textbf{2},
  033018 (2020).

\bibitem{Comaron_PhysRevResearch.2.022051}
P.~Comaron, V.~Shahnazaryan, W.~Brzezicki, T.~Hyart, and M.~Matuszewski,
  \enquote{Non-{H}ermitian topological end-mode lasing in polariton systems,}
  {\protect\JournalTitle{Phys. Rev. Research}} \textbf{2}, 022051 (2020).

\bibitem{Yamamoto_RMP}
H.~Deng, H.~Haug, and Y.~Yamamoto, \enquote{Exciton-polariton {B}ose-{E}instein
  condensation,} {\protect\JournalTitle{Rev. Mod. Phys.}} \textbf{82},
  1489--1537 (2010).

\bibitem{SAVONA1995733}
V.~Savona, L.~Andreani, P.~Schwendimann, and A.~Quattropani, \enquote{Quantum
  well excitons in semiconductor microcavities: Unified treatment of weak and
  strong coupling regimes,} {\protect\JournalTitle{Solid State Communications}}
  \textbf{93}, 733--739 (1995).

\bibitem{Schwendimann_2000}
G.~Rochat, C.~Ciuti, V.~Savona, C.~Piermarocchi, A.~Quattropani, and
  P.~Schwendimann, \enquote{Excitonic bloch equations for a two-dimensional
  system of interacting excitons,} {\protect\JournalTitle{Phys. Rev. B}}
  \textbf{61}, 13856--13862 (2000).

\bibitem{Ciuti_2003}
C.~Ciuti, P.~Schwendimann, and A.~Quattropani, \enquote{Theory of polariton
  parametric interactions in semiconductor microcavities,}
  {\protect\JournalTitle{Semiconductor Science and Technology}} \textbf{18},
  S279--S293 (2003).

\bibitem{Dominici2015}
L.~Dominici, M.~Petrov, M.~Matuszewski, D.~Ballarini, M.~De~Giorgi, D.~Colas,
  E.~Cancellieri, B.~Silva~Fern{\'a}ndez, A.~Bramati, G.~Gigli, A.~Kavokin,
  F.~Laussy, and D.~Sanvitto, \enquote{Real-space collapse of a polariton
  condensate,} {\protect\JournalTitle{Nature Communications}} \textbf{6}, 8993
  (2015).

\bibitem{Hofling_Dark}
D.~Schmidt, B.~Berger, M.~Kahlert, M.~Bayer, C.~Schneider, S.~H\"ofling, E.~S.
  Sedov, A.~V. Kavokin, and M.~A\ss{}mann, \enquote{Tracking dark excitons with
  exciton polaritons in semiconductor microcavities,}
  {\protect\JournalTitle{Phys. Rev. Lett.}} \textbf{122}, 047403 (2019).

\bibitem{Krizhanovskii_DarkSolitons}
P.~M. Walker, L.~Tinkler, B.~Royall, D.~V. Skryabin, I.~Farrer, D.~A. Ritchie,
  M.~S. Skolnick, and D.~N. Krizhanovskii, \enquote{Dark solitons in high
  velocity waveguide polariton fluids,} {\protect\JournalTitle{Phys. Rev.
  Lett.}} \textbf{119}, 097403 (2017).

\bibitem{Bobrovska_Adiabatic}
N.~Bobrovska and M.~Matuszewski, \enquote{Adiabatic approximation and
  fluctuations in exciton-polariton condensates,} {\protect\JournalTitle{Phys.
  Rev. B}} \textbf{92}, 035311 (2015).

\bibitem{Deveaud_Bistability}
M.~Wouters, T.~K. Para\"{\i}so, Y.~L\'eger, R.~Cerna, F.~Morier-Genoud, M.~T.
  Portella-Oberli, and B.~Deveaud-Pl\'edran, \enquote{Influence of a
  nonradiative reservoir on polariton spin multistability,}
  {\protect\JournalTitle{Phys. Rev. B}} \textbf{87}, 045303 (2013).

\bibitem{Adachi_PhysE2001}
S.~Adachi, T.~Tsuchiya, H.~Mino, S.~Takeyama, G.~Karczewski, T.~Wojitowicz, and
  J.~Kossut, \enquote{Dynamical spin properties of exciton and biexciton in
  {CdMnTe}/{CdTe}/{CdMgTe} single quantum well,} {\protect\JournalTitle{Physica
  E}} \textbf{10}, 305--309 (2001).

\bibitem{Seredynski_2018}
B.~Seredy\ifmmode~\acute{n}\else \'{n}\fi{}ski, M.~Kr\'ol, P.~Starzyk,
  R.~Mirek, M.~\ifmmode~\acute{S}\else \'{S}\fi{}ciesiek, K.~Sobczak,
  J.~Borysiuk, D.~Stephan, J.-G. Rousset, J.~Szczytko,
  B.~Pi\ifmmode~\mbox{\k{e}}\else \k{e}\fi{}tka, and W.~Pacuski,
  \enquote{{(Cd,Zn,Mg)Te}-based microcavity on {MgTe} sacrificial buffer:
  Growth, lift-off, and transmission studies of polaritons,}
  {\protect\JournalTitle{Phys. Rev. Materials}} \textbf{2}, 043406 (2018).

\bibitem{Furman_2023}
M.~Furman, A.~Opala, M.~Kr\'ol, K.~Tyszka, R.~Mirek, M.~Muszy\'nski,
  B.~Seredy\'nski, W.~Pacuski, J.~Szczytko, M.~Matuszewski, and B.~Pi\k{e}tka,
  \enquote{{Inverted optical bistability and optical limiting in coherently
  driven exciton--polaritons},} {\protect\JournalTitle{APL Photonics}}
  \textbf{8}, 046105 (2023).

\end{thebibliography}

\end{document}